# Infrastructure to Vehicle real Time Secured Communication

Ms.Smita Narendra Pathak [1], Prof.Urmila Shrawankar [2]

[1] Department of computer science and Engineering
G.H. Raisoni college of Engineering Nagpur, India, smita_27dec@yahoo.co.in
[2] Department of computer science and Engineering
G.H. Raisoni college of Engineering Nagpur, India, urmilas@rediffmail.com

**Abstract.** Among civilian communication systems, vehicular networks emerge as one of the most is convincing and yet most challenging instantiations of the mobile ad hoc networking technology. Towards the deployment of vehicular communication systems, security and privacy are critical factors and significant challenges to be met. This Vehicular communication (VC) system has the potential to improve road safety and driving comfort. Nevertheless, securing the operation is a prerequisite for deployment so in this paper we are focusing on real time experimental design of infrastructure to vehicle communication. We outline how VANET will be a better option than GPS technology. We also try to discuss IP address passing using DHCP in the network and the security issues.

## 1. Introduction

Vehicular networks are emerging as a new research area in mobile networking, as wireless ad hoc communication can enable equipped vehicles to exchange safety, transportation efficiency, and other information. Both academia and car manufacturers are progressively paying more and more attention to Vehicular Communications (VC), that allow vehicles to connect to each other and with the roadside infrastructure to form a vehicular Ad-hoc Network (VANET). The nodes of a VANET are commonly divided in two categories: On-Board Units (OBU), that are radio devices installed on vehicles, and Road Side Units (RSU), that constitute the network infrastructure. RSUs are placed along the roadside and are controlled by a network operator. An example of VC is IEEE 802.11p, a protocol belonging to the IEEE 802.11 family. The expected development of VANETs will involve millions of vehicles worldwide, making that network the most extended form of mobile ad-hoc networks [9]. Hence, VANETs introduce a number of challenges to the research community like high node mobility, nodes heterogeneity, and several security issues.

Before considering extending Internet legacy services to in-motion users, it is worth studying experimentally the 802.11 technology capabilities to support such applications while on the move. It is well known that TCP/IP-based applications do not work well under conditions where network connectivity is transient and where link quality is highly variable. The performance of TCP in wireless network scenarios has been well-studied [1]. Throughout many research efforts have attempted to mitigate these problems. Our motivation is to understand the factors limiting the performance of wireless networks for in-motion users using realistic applications. In other words, how well can a user moving past an access point, e.g., in a car or bus, make use of these transient opportunities to perform such tasks as browse the web, send and receive email messages, or conduct file transfers.

In this paper, we investigate the usability of providing wifi network connectivity to mobile users in vehicles.

The rest of the paper is organized as follows. First, we outline design setup of our real time experiment in section 2, and then we describe security requirement in section 3. Section 4 there is results obtained .Section 5 concludes the paper and at last there are the references.

## 2. Experimental setup

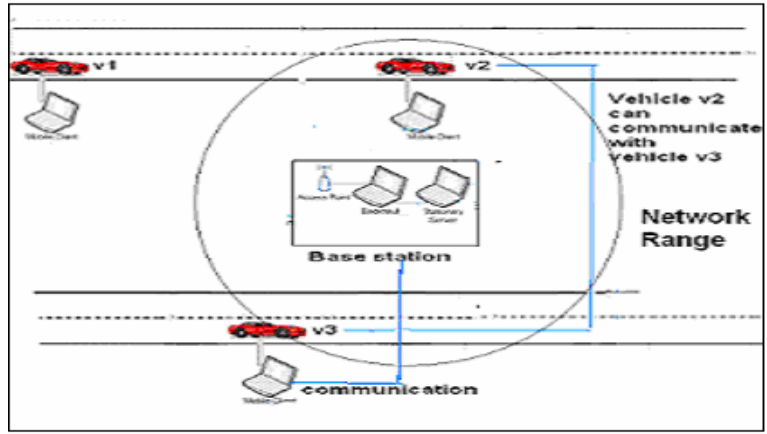

Fig.1 Experimental setup of V2V & I2V architecture

The hardware configuration of the client requires an IBM ThinkPad T41 with integrated Intel Pro/Wireless.2915ABG wireless adaptor.

The client communicates with a Linksys WAP55AG access point operating on channel 1 in 802.11b mode only. The APs WAN port is wired to a laptop modeling the backhaul network. That laptop featured an IBM Thinkpad T30 with one built-in ethernet adaptor and one PCMCIA ethernet adaptor. The T30's other ethernet adaptor is connected to the server, an IBM Thinkpad T42. Both the client and server are also equipped with a PCMCIA Netgear WAG511 802.11 wireless adaptor used for network monitoring only. We considered using DHCP, but Ott and Kutscher's analysis [2] showed highly variable performance due to slow retransmission timers. Therefore, we decided to use preset IP addresses for all machines. The hardware above was chosen to be typical of that used by real users. No external high-gain antennae were used on the clients or the access point, and default link layer parameters were always used.

The software configuration is as follows. All machines used the XP 2005 as a operating system and we will be configuring the systems by using VB.Net s/w.

### 2.1. Working concept of the setup
The above Fig. 1 shows the architectural setup of VANET (V2V and I2V) communication, in which server acting as a base station which can be kept at any infrastructural location and configured such that, if the vehicle is within the range of base station then it will starts communication with the base station where the base station displays its License number, speed and vehicle name, this information will be displayed by the base station for every vehicle coming in its communication rage. For our setup we set the rage of communication as the range of access point hence vehicle V1 can not established communication with base station as it is out of range of communication on the other hand vehicles v2 and v3 can established communication with base station as well as they can communicate any information within themselves. This concept will support both infrastructures to vehicle communication as well as vehicle to vehicle communication setup.

## 3. Security Requirements
The unique features of VC are a double-edged sword: a rich set of tools are offered to drivers and authorities but a formidable set of abuses and attacks becomes possible. Consider, for example, nodes that 'contaminate' large portions of the vehicular network with false information: a single vehicle can transmit false hazard warnings (e.g., ice formation on the pavement), which can then be taken up by all vehicles in both traffic streams. Or, similarly, a vehicle that meaningfully modifies messages of other vehicles. Or even a vehicle that forges messages in order to masquerade an emergency vehicle to mislead other vehicles to slow down and yield.These simple examples of exploits indicate that under all circumstances vehicular communications must be secured. In fact, it is possible that vehicles and their sensing, processing, and communication platforms are compromised. Worse even, any wireless enabled device that runs a rogue version of the vehicular communication protocol stack poses a threat both to the vehicular network and the transportation system operation. Hence, the security of vehicular networks is indispensable; otherwise these systems could make anti-social and criminal behavior easier, in ways that would actually jeopardize the benefits of their deployment. The problem at hand is to secure the operation of vehicular communication systems, that is, design protocols that mitigate attacks and thwart to the greatest possible extent deviations from the implemented protocols. Securing vehicular communications is a hard problem, with a broad range of challenges to be addressed. Different aspects warrant distinct protocols, and thus per (type of) protocol specifications. Instead of such specifications, we provide next an outline of general security requirements. They are rather stand-alone requirements and can be viewed as building blocks towards more complex specifications.

- **Message Authentication and Integrity:** Messages must be protected from any alteration and the receiver of a message must corroborate the sender of the message. Integrity, however, does not necessarily imply identification of the sender of the message.
- **Message Non-Repudiation:** The sender of a message cannot deny having sent a message.
- **Entity Authentication**: The receiver is not only ensured that the sender generated a message but in addition has evidence of the liveness of the sender. A received unmodified message was generated within an interval $[t − \tau, t]$, with t the current time at the receiver and $\tau > 0$ a sufficiently small positive value.
- **Access Control:** Access to specific services provided by the infrastructure nodes, or other nodes, is determined locally by policies. As discussed further in Sec. VI, access to network and messages is mandated by default open to all nodes. This, however, does not preclude the need for fine-grained policies for all other purposes, as well as the assignment of distinct roles to different types of nodes. As part of access control,
- **authorization:** establishes what each node is allowed to do in the network, e.g., which types of messages it can insert in the network, or more generally the protocols it is allowed toexecute
- **Message Confidentiality**: The content of a message is kept secret from those nodes that are not authorized to access it.
- **Privacy and Anonymity:** Vehicular communication systems should not disclose or allow inferences on the personal and private information of their users. This being a very general statement and a requirement within the broader area of information hiding, we state a narrower requirement within the vehicular network context: anonymity. We require anonymity for the actions (e.g., messages,transactions) of the vehicular network entities, which we denote as nodes, with respect to a set of observers. At minimum, any of the observers should not be able to learn if a node performed or will perform in the future a specific action, assuming that the node performs the action. Such a definition does not, however, guarantee that it is impossible for the observer to infer, with relatively high probability, the identity of the node that performs the action in question. To prevent such inferences, stronger anonymity requirements would be necessary: nodes should be almost equally likely to have performed an action, or have strong probabilistic anonymity, with the probabilities, as far an observer is concerned, being equal for any node or, without considering probabilities, require full anonymity.The definition of anonymity depends on what is the set of the VC system entities. or, in fact, whether entities are partitioned into a number of subsets, for administrative reasons. This implies that the anonymity requirement needs to be modified accordingly.

Availability Protocols and services should remain operational even in the presence of faults, malicious or benign. This implies not only secure but also fault-tolerant designs, resilience to resource depletion attacks, as well as self-stable protocols, which resume their normal operation after the 'removal' of the faulty participants.

Liability Identification Users of vehicles are liable for their deliberate or accidental actions that disrupt the operation of other nodes, or the transportation system. The vehicular network should provide information that identifies or assists the attribution of liability. This is a requirement that largely follows from the current practice in transportation systems. However, liability identification implies that anonymity would need to be paired with the option to learn or essentially recover the node's identity if necessary.

## 4. Implementation Results

As shown in fig 1.the first result is of authentication, where we need to provide a user name and password for starting our server. As the server starts , list of clients connected is displayed along with their IP address, identification name and timestamp. Client also stores the information about how many clients are connected at server along with their names and time stamps as shown in fig2Message can be send by the client to server or by server to client about any of the road hazards. The security is implemented as equal to WPA2 (wifi protected access) security. The message can be encrypted by using AES Algorithm as shown in fig.3

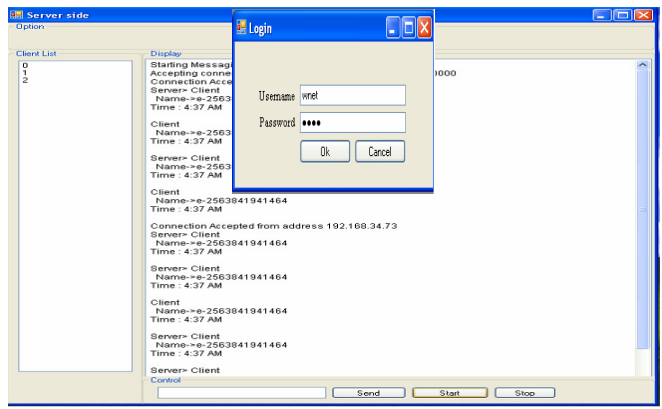

Fig:2 Server side configuration

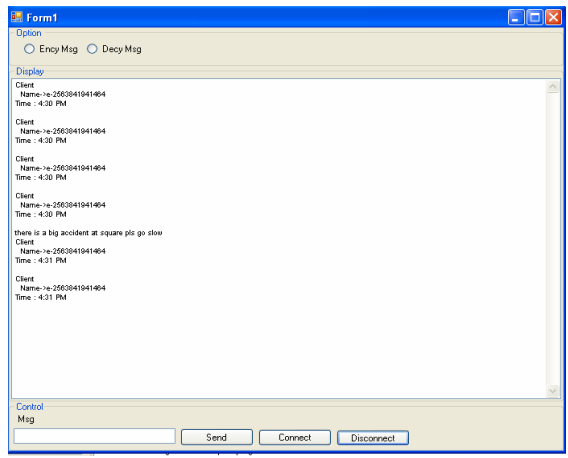

Fig:2 Client side configuration

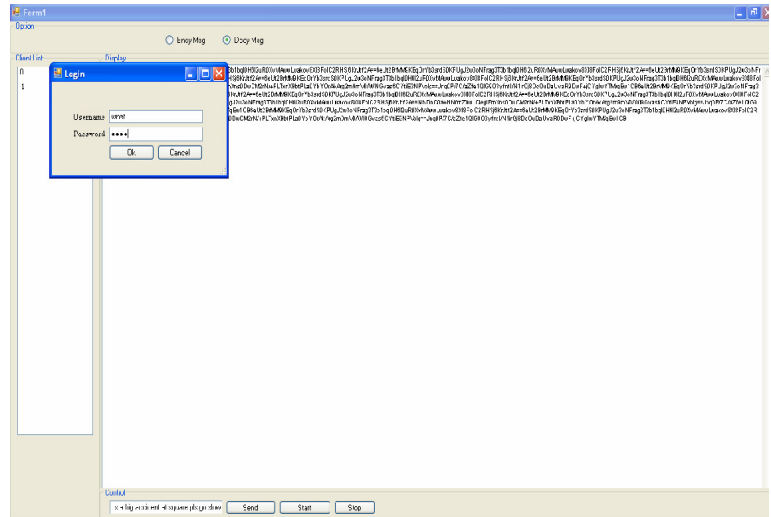

## 5. Conclusion:

According to the results obtained, if the vehicle is coming into the wi-fi range then it is identified at the base station. The message can be passed by the base station to client as well as client to the base station .Hence vehicle can be pre intimated of road conditions or hazards so that the accidents can be avoided. We try to implement the authentication and confidentiality security issue in wireless vehicle to infrastructure communication by using cryptography with the minimum requirements.

In the future the multimedia files can be transferred for the communication and video files can also be used for the more clear communication purpose.